\documentclass[preprintnumbers,amsmath,amssymb,prb,
twocolumn,superscriptaddress]{revtex4}

\usepackage{graphicx}   
\usepackage{dcolumn}    
\usepackage{bm}         

\begin{document}

\title{Probing the singlet-triplet splitting in double quantum dots: \\
implications of the ac field amplitude}

\author{G. Giavaras}
\affiliation{Faculty of Pure and Applied Sciences, University of
Tsukuba, Tsukuba 305-8571, Japan}

\author{Yasuhiro Tokura}
\affiliation{Faculty of Pure and Applied Sciences, University of
Tsukuba, Tsukuba 305-8571, Japan} \affiliation{Tsukuba Research
Center for Energy Materials Science (TREMS), Tsukuba 305-8571,
Japan}


\begin{abstract}
We consider a double quantum dot whose energy detuning is
controlled by an ac electric field. We demonstrate an energy
configuration for which the ac-induced current flowing through the
double dot directly probes the spin-orbit anticrossing point for
small ac field amplitudes. On the contrary, as the ac amplitude
increases a current antiresonance is formed, and the direct
information about the spin-orbit interaction is lost. This result
indicates that a large ac amplitude is not necessarily
advantageous for the spectroscopy of spin-orbit coupled two-spin
states. Moreover, we investigate the ac-induced current peaks
versus the ac amplitude and show a current suppression when the ac
field forms spin blocked states. This effect gives rise to a
characteristic pattern for the current which can be controlled at
will by tuning the ac amplitude. Our results can be explored by
performing electronic transport measurements in the spin blockade
regime.
\end{abstract}

\maketitle

\section{Introduction}

Various spin qubit proposals in semiconductor materials make use
of electron spins trapped in quantum dot systems.~\cite{hanson07,
zwanenburg13} Spin-orbit coupled spins defined in double dots at a
constant magnetic field, can be manipulated electrically using an
ac electric field.~\cite{perge12, ono17} One clear signature of
the spin-orbit interaction (SOI) is the formation of
singlet-triplet anticrossing points in the two-electron energy
spectrum. The magnitude of the energy gap at the anticrossing
point is an important energy scale because it gives information
about the strength and the direction of the SOI.~\cite{anticros1,
anticros2, nowak, takahashi, kanai, stano, fasth} Usually a large
gap is the result of strong SOI. Transport spectroscopy of the
two-electron energy spectrum can be performed by measuring the
electrical current through the double dot in the presence of an ac
electric field.~\cite{perge12, ono17} Current peaks arise when the
appropriate resonant condition is satisfied,~\cite{giavaras19} and
information about the SOI can be extracted provided the ac-induced
current peaks are well-formed in the vicinity of the SOI
anticrossing point.

The applied ac field is characterized by the ac frequency and
amplitude. The range of the ac frequency is dictated by the energy
configuration of the two-electron eigenstates in the double dot
and the relevant energy splitting. Therefore, the ac-induced peaks
can be controlled by the ac amplitude only, provided the ac
amplitude can be tuned by the applied voltages to gate electrodes.

In this work, we consider a double dot (DD) in the spin blockade
regime,~\cite{ono00} and focus on experimentally accessible DD
energy configurations, where singlet and triplet energy levels
anticross. In particular, the focus is on two SOI-coupled
singlet-triplet states forming an anticrossing point, and a third
state with triplet character. We assume that an ac field
periodically changes the energy detuning of the DD, in the same
way as in the experiments,~\cite{perge12, ono17} and investigate
possible implications of the magnitude of the ac amplitude in the
ac-induced current peaks. We show that the current peaks allow for
transport spectroscopy of the SOI anticrossing point only in a
specific ac amplitude range, which is related to the strength of
the SOI. When the ac amplitude is large the energy gap of the
anticrossing can no longer be probed accurately, and instead an
``antiresonance'' is formed, where typically the ac current is
suppressed. As a consequence, a large ac amplitude is not
necessarily advantageous for spectroscopy, especially when the
presence of the SOI is directly inferred by the current
characteristics versus the ac frequency and magnetic field.
Furthermore, we study the dependence of the ac-induced current on
the energy detuning as well as the ac field amplitude, and
identify a rather general pattern of high- and low-current
regions. These regions stem from the formation of ac-induced spin
blocked states, and thus can be controlled at will by tuning the
ac amplitude.

In the next section, the double quantum dot model and the
electronic transport model are presented. In Sec.~\ref{transport}
the ac-induced transport characteristics for different ac field
frequencies and amplitudes are studied. The basic conclusions of
this work are summarized in Sec.~\ref{conclusion}.

\section{Physical model}\label{model}

\subsection{Double dot Hamiltonian}

In this work, we consider two serially tunnel-coupled quantum dots
in the spin blockade regime~\cite{ono00}, and assume the dot
charging energy to be much larger than the inter-dot tunnel
coupling. The quantum dot 1 (dot 2) is coupled to the left (right)
metallic lead, therefore under an appropriate bias voltage current
can flow through the system which is sensitive to spin
correlations. We assume that each dot is characterised by a single
orbital level (on-site energy), and dot 2 is lower in energy so
that a single spin is localised in dot 2, and the spin blockade
regime can be realised.~\cite{hanson07, zwanenburg13, ono00} In
this regime the electronic transport through the DD system follows
the charge cycle~\cite{hanson07, zwanenburg13, ono00}: $(0, 1)
\rightarrow (1, 1) \rightarrow (0, 2) \rightarrow (0, 1)$, where
the notation $(n, m)$ indicates $n$ electrons on dot 1 and $m$
electrons on dot 2. The relevant two-electron states are the (1,
1) triplet states $|T_{+}\rangle$, $|T_{-}\rangle$,
$|T_{0}\rangle$, the (1, 1) singlet state $|S_{11}\rangle$, and
the (0, 2) singlet state $|S_{02}\rangle$. The (2, 0) singlet
state $|S_{20}\rangle$ is much higher in energy and to a very good
approximation, can be ignored without affecting the
physics.~\cite{hanson07, zwanenburg13}

In the Appendix, we show that for two electrons and in the basis
$|S_{11}\rangle$, $|T_{+}\rangle$, $|S_{02}\rangle$,
$|T_{-}\rangle$, $|T_{0}\rangle$ the DD Hamiltonian is
\begin{equation}\label{hamilton}
H_{\mathrm{DD}} =\left(%
\begin{array}{ccccc}
  0 & 0 & -\sqrt{2}t_{\mathrm{c}} & 0 & \Delta^{-} \\
  0 & -\Delta^{+} & -t_{\mathrm{so}} & 0 & 0 \\
  -\sqrt{2}t_{\mathrm{c}} & -t_{\mathrm{so}} & \delta & -t_{\mathrm{so}} & 0 \\
  0 & 0 &  -t_{\mathrm{so}} & \Delta^{+} &0\\
  \Delta^{-} & 0 & 0 & 0 & 0 \\
\end{array}%
\right).
\end{equation}
The Zeeman term on dot $i$ ($i=1$, 2) is given by $\Delta_{i}=g_i
\mu_{\mathrm{B}} B$, where $B$ is the external magnetic field and
$g_{i}$ is the $g$-factor with
$\Delta^{\pm}=(\Delta_{1}\pm\Delta_{2})/2$. The parameter
$t_{\mathrm{c}}$ is the inter-dot tunnel coupling which conserves
spin, $t_{\mathrm{so}}$ is the spin-flip tunnel coupling due to
the SOI, and $\delta$ is the energy detuning. Some experimental
studies~\cite{perge12, ono17, wang, chorley11} on double quantum
dots conclude that for the tunnel couplings
$t_{\mathrm{so}}<t_{\mathrm{c}}$, and in this work we satisfy this
condition. The one-electron states (0, 1) consist of the spin-up
$|0, \uparrow\rangle$, and spin-down $|0, \downarrow\rangle$
configurations which are Zeeman-split due to the magnetic field
$B$. The one-electron states (1, 0) can usually be ignored in the
spin blockade regime provided the dots are weakly
coupled.~\cite{hanson07, zwanenburg13}

We assume that an ac electric field periodically modulates the
on-site orbital energy of dot 2, relative to dot 1. In this case,
we can consider the energy of the (1, 1) states to be unaffected
by the ac field, and the energy of the (0, 2) state to be time
dependent. Thus, according to Hamiltonian Eq.~(\ref{hamilton}) the
energy detuning in this work is considered to be time periodic
\begin{equation}\label{detuning}
\delta(t) = - \varepsilon+A\cos(2\pi f t),
\end{equation}
where $A$, $f$ are the amplitude and frequency of the ac field
respectively. In semiconductor quantum dots the value of
$\varepsilon$ is controlled by applying appropriate gate voltages
~\cite{hanson07, zwanenburg13, bertrand} and the values of $A$,
$f$ are tunable by electrical pulses.~\cite{perge12, ono17,
bertrand, pulses1}

For all the calculations the inter-dot tunnel coupling is taken to
be $t_{\mathrm{c}}=13$ $\mu$eV, in agreement with experimentally
reported values.~\cite{hanson07, zwanenburg13} The $g$-factors of
the two dots are taken to be $g_1=7$, and $g_{2}=7.5$. These
absolute values are within the range of the $g$-factors reported
for InAs systems.~\cite{smith87} The $g$-factor difference of
about 8$\%$ is consistent with that found in double quantum dots,
and could be the result of the SOI, and/or the asymmetric double
dot confining potential. Even larger $g$-factor differences have
been reported. For instance, in Ref.~\onlinecite{berg} the
absolute $g$-factor difference in an InSb double quantum dot with
strong SOI was measured to be as large as 12; e.g. over 20$\%$
difference.

\begin{figure}
\includegraphics[width=10.5cm, angle=270]{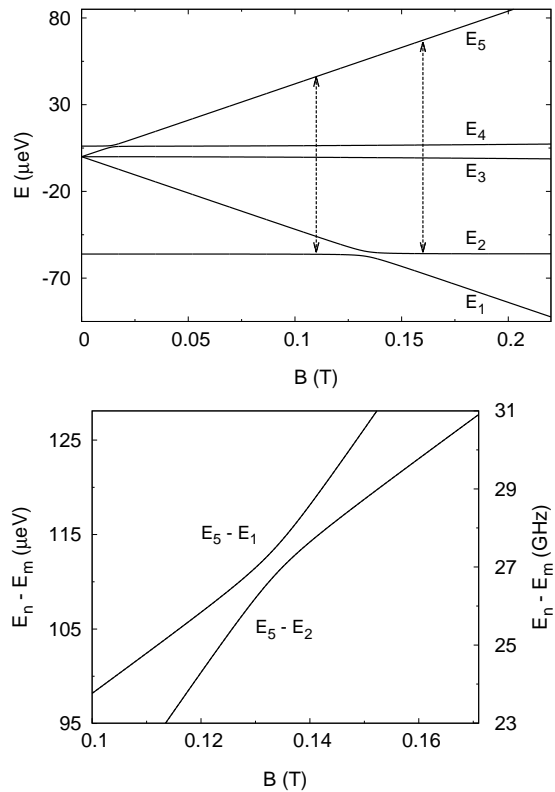}\\
\caption{The upper frame shows the two-electron eigenenergies as a
function of magnetic field. The two vertical arrows indicate
possible transitions that can be induced by the ac electric field,
which periodically changes the energy detuning of the double
quantum dot. The lower frame shows the energy splitting
$E_{5}-E_{1}$ and $E_{5}-E_{2}$ as a function of magnetic
field.}\label{energy}
\end{figure}

The eigenstates of the DD Hamiltonian $H_{\mathrm{DD}}$ for $A=0$,
are denoted by $|\psi_{n}\rangle$, $n=1$, 2,...5, and are ordered
in increasing eigenenergy. We refer to $|\psi_{n}\rangle$ as
singlet or triplet states, though $|\psi_{n}\rangle$ consist of
both singlet and triplet components due to the SOI and the
$g$-factor difference in the two dots. Thus, the spin blockade can
be lifted and the ac field can induce singlet-triplet transitions.
The corresponding DD eigenenergies $E_{n}$, $n=1$, 2,...5, versus
the magnetic field are shown in the upper frame of Fig.~1, for
$t_{\mathrm{so}}=1.5$ $\mu$eV and $\varepsilon=50$ $\mu$eV. In
this work, we are interested in the region of the SOI induced
anticrossing point which is formed at $B\approx0.134$ T, and the
corresponding gap is about 0.7 GHz. The ac field induced
transitions of interest are between the state $|\psi_{5}\rangle$
which has triplet character, and the two SOI-coupled
singlet-triplet states $|\psi_{1}\rangle$, $|\psi_{2}\rangle$
forming the anticrossing point. In particular, the two vertical
arrows shown in the upper frame of Fig.~1 specify the ac field
induced transitions which are under investigation, e.g.,
$hf\approx E_{5}-E_{1}$ and $hf\approx E_{5}-E_{2}$, where $h$ is
Planck's constant, and $E_{5}-E_{1}$ and $E_{5}-E_{2}$ are shown
in the lower frame of Fig.~1. This loose view does not imply that
the other eigenstates, not directly involved in the transitions,
are in general not relevant to the ac field induced dynamics. The
transitions between the singlet-triplet states $|\psi_{1}\rangle$
and $|\psi_{2}\rangle$ can also give information about the
anticrossing point,~\cite{ono17} but these transitions are not
considered in the present work.

When the ac field modulates the potential profile of the DD
leading to a time dependent energy detuning as described in
Eq.~(\ref{detuning}), the inter-dot potential barrier may also
acquire a (small) time dependence. This in turn means that in our
model the inter-dot tunnel coupling can be time dependent, and can
therefore result in singlet-triplet transitions.~\cite{giavaras19}
Here, we assume that the time dependence of the tunnel coupling is
negligible and can be safely ignored.

\subsection{Master equation formalism}

In this subsection we briefly describe the basic features of the
quantum transport model which is based on a Floquet-Markov master
equation.~\cite{flqmas1, flqmas2} The dot 1 (dot 2) is
tunnel-coupled to the left (right) lead, and under an appropriate
bias voltage in the spin blockade regime electrons flow through
the system.~\cite{ono00} The electrons in the two leads are
assumed to be non-interacting and described by the Hamiltonian
\begin{equation}
H_{\mathrm{e}}=\sum_{\ell, k,\sigma}\epsilon_{\ell
k}d^{\dagger}_{\ell k\sigma} d_{\ell k\sigma}.
\end{equation}
The operator $d^{\dagger}_{\ell k \sigma}$ ($d_{\ell k\sigma}$)
creates (annihilates) an electron in the lead $\ell=\{$L, R$\}$,
with momentum $k$, spin $\sigma$, and energy $\epsilon_{\ell k}$.
Electron tunnelling between the two leads and the DD is described
by the Hamiltonian
\begin{equation}
H_{\mathrm{T}} = t_{\mathrm{T}}\sum_{
k,\sigma}(c_{1\sigma}^{\dagger}d_{\mathrm{L}
k\sigma}+c_{2\sigma}^{\dagger}d_{\mathrm{R} k\sigma})
+\text{H.c.},
\end{equation}
Here, $c^{\dagger}_{i\sigma}$ is the electron creation operator on
dot $i$ with spin $\sigma$, and $t_{\mathrm{T}}$ is the dot-lead
coupling constant.

We are interested in finding the density matrix $\rho(t)$ of the
DD, and because the DD Hamiltonian is time periodic
$H_{\mathrm{DD}}(t)=H_{\mathrm{DD}}(t+T)$, with $T=1/f$, we choose
to express the density matrix $\rho(t)$ in the Floquet modes basis
$|u(t)\rangle$. This choice significantly simplifies the master
equation of motion of $\rho(t)$, because the steady-state can be
extracted without performing a numerical time integration which is
usually time-consuming. The Floquet modes are periodic,
$|u(t)\rangle=|u(t+T)\rangle$, and satisfy the Floquet eigenvalue
problem,
\begin{equation}
\left( H_{\mathrm{DD}}(t) - i\hbar \frac{\partial}{\partial t}
\right) |u_j(t)\rangle = \kappa_j |u_{j}(t)\rangle,
\end{equation}
where $\kappa_j$ are the corresponding Floquet energies. The
Floquet modes are expanded in the singlet-triplet basis
\begin{equation}
|u_j(t)\rangle =  \sum^{5}_{n=1} b_{j,n}(t)|\text{ST}_{n}\rangle,
\end{equation}
with the coefficients $b_{j,n}(t)=b_{j,n}(t+T)$, and
$|\text{ST}_{n}\rangle$ are the singlet-triplet basis vectors.
Both $H_{\mathrm{DD}}(t)$ and $b_{j,n}(t)$ are expanded in a
Fourier series and the resulting eigenvalue problem is solved
numerically. The Floquet energy spectrum consists of identical
energy zones of width $hf$, and inspection of one of the zones
provides information on the resonant condition(s) as the ac
amplitude increases.~\cite{shirley} In contrast, the bare
eigenenergies of the time independent part of $H_{\mathrm{DD}}$
fail to predict the well-known frequency shifts in the context of
the Bloch-Siegert theory.~\cite{bloch}

The equation of motion of the density matrix $\rho(t)$ of the DD
takes into account sequential electron tunnelling from the leads
into the DD and vice versa, with a change in the electron number
by $\pm 1$. Using for the matrix elements of $\rho(t)$ the
notation $\rho_{nj}(t) = \langle u_{n}(t) | \rho(t) | u_{j}
(t)\rangle$, and for the Floquet energies $\kappa_{nj} =
\kappa_{n} - \kappa_{j}$ the equation of motion can be written as
follows
\begin{equation}\label{master}
\begin{split}
&\left(  \frac{\partial}{\partial t} + \frac{i}{\hbar}
\kappa_{nj} \right)\rho_{nj}(t) =\\
&\sum_{m, l}\lbrace - \rho_{lj}(t) X_{nm;lm}(t) -
\rho_{nm}(t) Q_{lj;lm}(t)\\
&+ \rho_{ml}(t)[ Q_{nm;jl}(t) + X_{lj;mn}(t)]
\rbrace.\\
\end{split}
\end{equation}
The tensors $X$, $Q$ define the transition rates which determine
the dot-lead tunnelling. In the steady-state,
$\rho(t)=\rho_{\mathrm{st}}(t)$, and we assume that
$\rho_{\mathrm{st}}(t)$ is periodic with the same period as that
of the ac field. For the regime of parameters in this work we can
further assume that to a good approximation
$\rho_{\mathrm{st}}(t)$ is equal to its zero frequency Fourier
component. Then, $\rho_{\mathrm{st}}$ becomes approximately time
independent, and this can also be assumed to be the case for $X$
and $Q$. If we consider the interaction of dot 2 with the right
lead and, for simplicity, the spin-up only contribution, then
\begin{equation}
\begin{split}
&X_{in;lj}
=\Gamma\sum^{\infty}_{L=-\infty}\lbrace[c_{2\uparrow}(L)]_{in}
[c_{2\uparrow}(L)]^*_{lj}f_{\mathrm{R}}(\kappa_{jl}-L\hbar\omega)\\
& + [c_{2\uparrow}(L)]^{*}_{ni}
[c_{2\uparrow}(L)]_{jl}f^{-}_{\mathrm{R}}(-\kappa_{jl}-L\hbar\omega)\rbrace,\\
\end{split}
\end{equation}
with the matrix elements
\begin{equation}
[c_{2\uparrow}(M)]_{nm}=\frac{1}{T}\int^{T}_{0} e^{-i M \omega t}
\langle u_{n}(t) |c_{2\uparrow}|u_{m}(t)\rangle dt,
\end{equation}
and the cyclic frequency $\omega = 2 \pi f$. The density of states
$D$ of the right lead is taken to be energy independent leading to
the constant dot-lead tunnelling rate
$\Gamma=2\pi|t_{\mathrm{T}}|^2 D/\hbar$. The Fermi function of the
right lead is $f_{\mathrm{R}}$ with
$f^{-}_{\mathrm{R}}=1-f_{\mathrm{R}}$, and $Q$ is found from $X$
by replacing $f_{\mathrm{R}} \rightarrow f^{-}_{\mathrm{R}}$,
$f^{-}_{\mathrm{R}} \rightarrow f_{\mathrm{R}}$. The matrix
elements involve one- and two-electron Floquet modes, but only for
the latter is a numerical computation needed. Moreover, in the
transition rate $X$ the same notations $|u_j(t)\rangle$ for the
Floquet modes, and $\kappa_j$ for the Floquet energies are
considered for both one and two electrons. The interaction of dot
1 with the left lead can be treated in the same way, and the
resulting equation of motion is solved numerically. Finally, the
current flowing through the right lead is given by the average of
the current operator $I = - e i [H_{\mathrm{T}},
N_{\mathrm{R}}]/\hbar$ where $N_{R}$ is the electron number
operator for the right lead, and $[H_{\mathrm{T}},
N_{\mathrm{R}}]= t_{\mathrm{T}}\sum_{ k,\sigma}
(c_{2\sigma}^{\dagger}d_{\mathrm{R} k\sigma} -
d^{\dagger}_{\mathrm{R} k\sigma} c_{2\sigma} )$.

\begin{figure}
\includegraphics[width=7.5cm, angle=0]{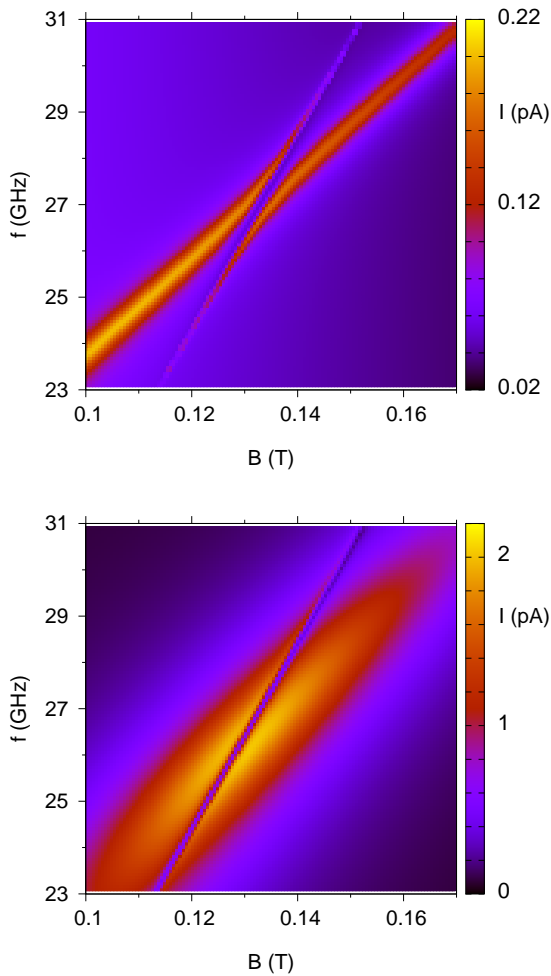}\\
\caption{Current as a function of ac frequency and magnetic field.
For the upper frame the ac amplitude is $A=5$ $\mu$eV, and for the
lower frame $A=100$ $\mu$eV.}\label{image}
\end{figure}

\section{ac-induced transport characteristics}\label{transport}

In this section the ac-induced current is computed for different
ac field frequencies and amplitudes, and the focus is on the two
transitions which are depicted schematically in the upper frame of
Fig.~1. The dot-lead tunnelling rate is $\Gamma=170$ MHz, and the
energy detuning is $\varepsilon=50$ $\mu$eV unless otherwise
specified.

\subsection{Current versus ac frequency}\label{curvsf}

Figure~\ref{image} shows the current as a function of the ac field
frequency and magnetic field for two different ac amplitudes $A$.
The frequency and magnetic field ranges are sensitive to the
energy detuning ($\varepsilon=50$ $\mu$eV). Larger values of
detuning require higher magnetic fields and ac frequencies, but ac
frequencies on the order of 50 GHz are within experimental
reach.~\cite{fujisawa1, fujisawa2} When $A=5$ $\mu$eV two curves
of high-current are formed. These curves can be attributed to the
two singlet-triplet resonant transitions depicted schematically in
Fig.~\ref{energy}. When the condition $hf \approx E_{5}-E_{1}$ or
$hf \approx E_{5}-E_{2}$ is satisfied an ac-induced current peak
is formed. The peak width is sensitive to the character of the
involved states, and the peak is broad when the singlet character
dominates over the triplet. For this reason the visibility of the
two curves of high current is enhanced near the anticrossing
point, i.e., $f\approx 27$ GHz and $B\approx 0.134$ T. Away from
the anticrossing point the SOI induced singlet-triplet coupling
weakens and the two curves acquire very different widths. The
reason is that for the transition between $|\psi_1\rangle$ and
$|\psi_5\rangle$ both involved states have triplet character,
whereas for the transition between $|\psi_2\rangle$ and
$|\psi_5\rangle$ the state $|\psi_2\rangle$ has singlet character.
In essence, for $A=5$ $\mu$eV the two curves of high current
map-out the singlet-triplet energy levels, and the SOI gap which
is about 0.7 GHz, can be directly extracted from the current plot.
This procedure has been demonstrated experimentally in different
types of double quantum dots.~\cite{perge12, ono17}

In contrast, when the ac amplitude is $A=100$ $\mu$eV, as shown in
Fig.~\ref{image}, the two curves of high current can no longer be
clearly distinguished. Moreover, when the condition $(g_1+g_2)
\mu_{\mathrm{B}} B = h f$ is satisfied an ``antiresonance'' is
formed, i.e., the ac-induced current is approximately equal to the
background current ($A=0$). The antiresonance is more pronounced
near the anticrossing point ($f\approx 27$ GHz, $B\approx 0.134$
T). At a fixed field $B$, the frequency $f$ at which the
antiresonance is formed, is not explicitly related to the ac
amplitude. However, we show below that when the ac amplitude
increases the ac induced current peaks start to overlap favoring
the observation of the antiresonance.

It has been demonstrated~\cite{ono17} that the ac-induced current
peaks vanish very near the anticrossing point, when the ac-induced
transitions involve the two eigenstates ($|\psi_1\rangle$ and
$|\psi_2\rangle$) which form the anticrossing. The results in
Fig.~\ref{image} demonstrate that the effect of the ac field can
be different when the transitions include a third eigenstate not
explicitly involved in the anticrossing. This is due to the large
population difference between the eigenstates. In particular, the
eigenstate $|\psi_5\rangle$ has triplet-like character, therefore
it is highly populated, whereas very near the anticrossing
$|\psi_1\rangle$ and $|\psi_2\rangle$ have almost identical
characters and are almost equally populated. As a result, the
effective transition rate between $|\psi_5\rangle$ and
$|\psi_1\rangle$ (or $|\psi_2\rangle$) is much higher compared to
the rate between $|\psi_1\rangle$ and $|\psi_2\rangle$. The peak
height is sensitive to the dot-lead tunnelling rate $\Gamma$, and
increasing $\Gamma$ at fixed ac amplitude $A$ tends to suppress
the peaks.

Based on the results in Fig.~\ref{image} we can conclude that the
singlet-triplet energy levels which anticross cannot be probed at
arbitrary ac amplitudes. This conclusion sets an important
constraint on the ac amplitude. In some transport
experiments,~\cite{perge12, ono17} the detection of a
singlet-triplet anticrossing gap with an ac field is a reliable
signature of the presence and strength of the SOI. However, the
results in Fig.~\ref{image} suggest that the SOI when combined
with an ac electric field can produce current characteristics
which do not explicitly reveal the anticrossing gap. Thus, the
detection of the SOI gap requires the appropriate choice of the ac
amplitude.

\begin{figure}
\includegraphics[width=5.7cm, angle=270]{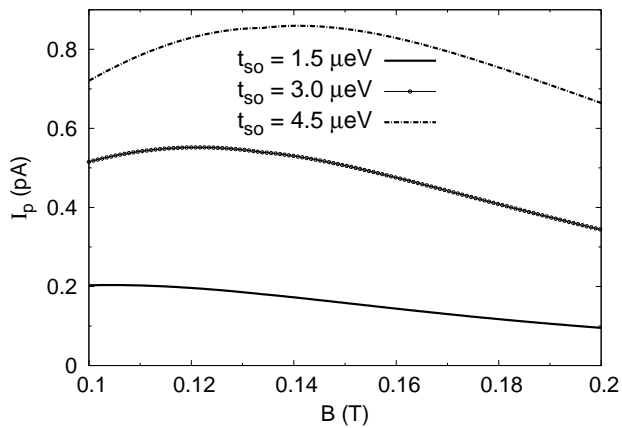}\\
\caption{Resonant current (peak height) as a function of magnetic
field for $A=5$ $\mu$eV, and different spin-orbit tunnel
couplings.}\label{reson}
\end{figure}

In Fig.~\ref{reson}, we tune the magnetic field in the range 0.1 T
$\le B \le 0.2$ T and plot, for each $B$ the resonant current,
i.e., the peak height. Here, we take~\cite{note1} $A=5$ $\mu$eV
and different SOI tunnel couplings $t_{\mathrm{so}}$. The peak
corresponds to the magnetic field dependent ac frequency, namely,
$f=(E_{5}-E_{i})/h$, where $E_{i}$ is the energy level of the
state with singlet character, and therefore $i=1$ or 2 (see also
Fig.~1). The peak height increases with $t_{\mathrm{so}}$ since
the singlet-triplet mixing increases leading to an enhanced
transition rate. The $B$ field at which the peak height is maximum
depends on $t_{\mathrm{so}}$, and can be different from the
anticrossing point ($B\approx 0.134$ T). This shows the overall
importance of the background populations (defined for $A=0$) of
the eigenstates, which are sensitive to $t_{\mathrm{so}}$ and also
to the $B$ field.~\cite{giavaras13} In this context, the
$g$-factor difference between the two dots affects the populations
by coupling $|T_0\rangle$ to singlet states, but the results in
Fig.~\ref{image} demonstrate that the SOI anticrossing point can
be probed even when the maximum current occurs away from the
anticrossing point.

\begin{figure}
\includegraphics[width=12.0cm, angle=270]{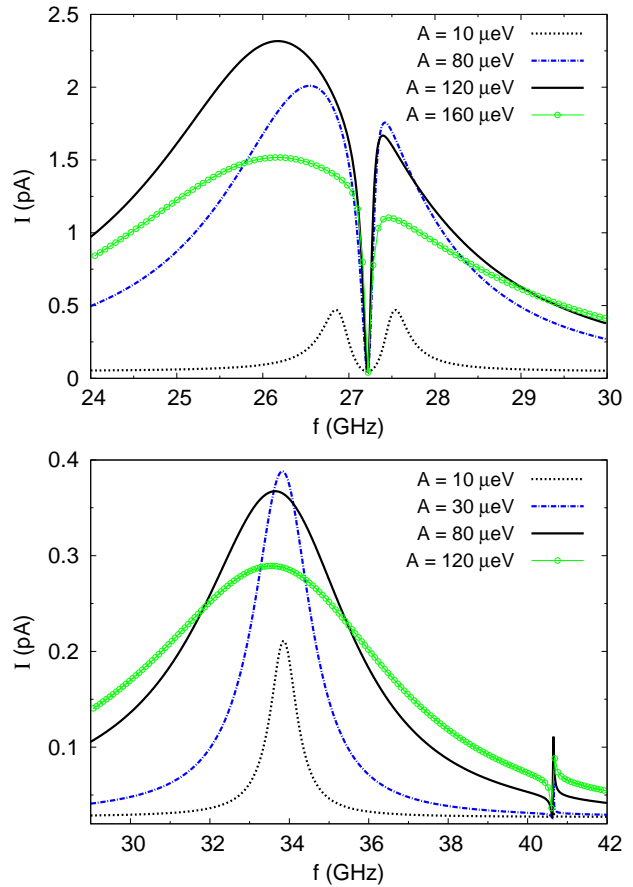}\\
\caption{Current as a function of ac frequency for different ac
amplitudes. For the upper frame the magnetic field is $B=0.134$ T,
and for the lower frame $B=0.2$ T.}\label{ivsf}
\end{figure}

To examine in more detail the pattern of the current, we plot in
Fig.~\ref{ivsf} the current as a function of the ac frequency $f$
for various ac amplitudes $A$. In this case we choose two fixed
values for the magnetic field; $B=0.134$ T which corresponds to
the anticrossing point, and $B=0.2$ T which is far from the
anticrossing point. When $A$ is small two peaks can be identified
that are centered at the resonant frequencies $f_{1}$ and $f_{2}$
where $h f_{i}\approx E_{5}-E_{i}$, $i=1$, 2. Consequently, the
corresponding singlet-triplet splitting is approximately given by
$h(f_2-f_1)$. Increasing $A$ results in broader peaks which
gradually start to overlap; this behaviour is more evident at
$B=0.134$ T. Provided $A$ is small enough such that the peaks have
negligible overlap, an approximate expression for the transition
rates in the coherent regime can be derived using a similar
methodology to that developed in Ref.~\onlinecite{giavaras19}. For
large $A$ Landau-Zener dynamics is relevant and one case for a
four-level quantum dot system has been studied
recently.~\cite{shev}

As seen in Fig.~\ref{ivsf} the peak height is in general different
for the two values of magnetic field. This is due to the fact that
the transition rates as well as the background populations ($A=0$)
of the eigenstates involved in the transitions are in general
magnetic field dependent, even for $g_{2} \approx g_{1}$. The peak
height also changes significantly with $A$. According to
Fig.~\ref{ivsf} the peak height increases with $A$ up to a maximum
value and then starts to decrease. For $B=0.2$ T the maximum
occurs at $A\approx 30$ $\mu$eV, and for $B=0.134$ T the maximum
occurs at $A\approx 120$ $\mu$eV. The dependence of the current on
the ac amplitude is examined below.

\begin{figure}
\includegraphics[width=10.5 cm, angle=270]{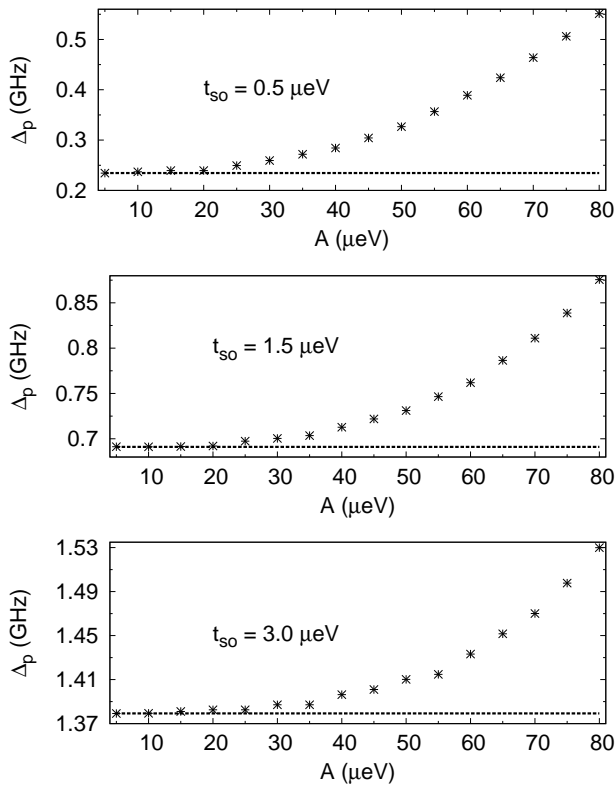}\\
\caption{The figure shows the distance $\Delta_{\mathrm{p}}$
between the two ac-induced current peaks which are centered at the
left and right of the antiresonance, for different ac amplitudes
$A$. The dotted line denotes the value of the SOI anticrossing gap
derived from the two-electron eigenenergies. The magnetic field
corresponds to the anticrossing point $B=0.134$ T, and the
spin-orbit tunnel coupling $t_{\mathrm{so}}$ is indicated in each
frame.}\label{ivstso}
\end{figure}

A current antiresonance has been theoretically predicted to arise
in a Coulomb blockaded DD in the presence of two microwave
fields,~\cite{brandes} and in a spin blockaded DD with a Zeeman
asymmetry which is driven by an oscillating magnetic
field.~\cite{giavaras10} A current antiresonance can also be
formed without a microwave irradiation.~\cite{sun} The important
conclusion of this section, i.e., the SOI anticrossing point
cannot be probed at arbitrary ac amplitudes, is independent of the
formation of the antiresonance, and the $g$-factor difference.

According to Fig.~\ref{ivsf} (upper frame), when the magnetic
field corresponds to the anticrossing point $B= 0.134$ T, the
ac-induced current peaks versus the ac frequency may be used to
estimate the values of the SOI gap, under the condition that the
ac amplitude $A$ is small. To quantify this condition we measure
the distance $\Delta_{\mathrm{p}}$ between the two current peaks
centered at the left and right of the antiresonance, and compare
$\Delta_{\mathrm{p}}$ with the exact value of the SOI gap derived
from the exact two-electron eigenenergies of the Hamiltonian
Eq.~(1) (for $A=0$). Figure~\ref{ivstso} shows the distance
$\Delta_{\mathrm{p}}$ between the peaks for different ac
amplitudes $A$, and three values for the tunnel coupling
$t_{\mathrm{so}}$. The exact value of the SOI gap is also
indicated. In all cases, when $A\lesssim 35$ $\mu$eV, the value of
the distance $\Delta_{\mathrm{p}}$ predicts the value of the exact
SOI gap with a small error. The relative error decreases with
$t_{\mathrm{so}}$ because the corresponding SOI gap increases. As
an example, for $A \approx 30$ $\mu$eV, the relative error is
about 11$\%$ for $t_{\mathrm{so}}=0.5$ $\mu$eV, whereas the
relative error is about 0.6$\%$ for $t_{\mathrm{so}}=3.0$ $\mu$eV.
An important aspect is that the peak width depends not only on the
ac amplitude but also on the strength of the SOI which in our
model is determined by the tunnel coupling $t_{\mathrm{so}}$, and
the energy detuning $\varepsilon$. The reason is that, for the
parameter range of this work, the ac-induced transition rates are
enhanced with $t_{\mathrm{so}}$, and the electrically driven
transitions we study here vanish when $t_{\mathrm{so}}=0$. The
results in Fig.~\ref{ivstso} also demonstrate that the error tends
to increase with $A$, since the two peaks start to overlap and
shift (Fig.~\ref{ivsf}), and as a consequence the value of
$\Delta_{\mathrm{p}}$ deviates from the exact SOI gap.

Even though a smaller ac amplitude can lead to a more accurate
estimation of the SOI gap, the dot-lead tunnelling rate $\Gamma$
sets another constraint on the ac amplitude $A$. A small $A$ can
give rise to coherent effects only when $\Gamma$ is small,
eventually inducing a small current which might be difficult to
measure. Measuring the ac-induced current peaks for only one value
of the ac amplitude may not be conclusive because the degree of
overlap of the current peaks cannot be inferred. Therefore, a more
efficient strategy to probe the SOI gap would be to tune the ac
amplitude and monitor the behaviour of the current peaks.

\subsection{Approximate Hamiltonian}\label{approx}

Some insight into the current characteristics can be obtained
within an approximate time independent Hamiltonian. It has been
shown that a single spin driven by an alternating magnetic field
displays resonances (single or multiphoton) when the two Floquet
energies anticross.~\cite{shirley} This property is general enough
and has been employed to predict the existence of resonances for
two coupled spins whose energy levels are time
dependent.~\cite{satanin, shevchenko} In this context, the
eigenenergies of the approximate time independent Hamiltonian
should also exhibit anticrossing points when a resonance occurs.
This remark is relevant when the starting point is the exact
Floquet Hamiltonian, as well as when deriving an approximate
Hamiltonian without directly employing the Floquet formalism.

To derive an approximate Hamiltonian we start with the time
dependent DD Hamiltonian Eq.~(1), and apply a unitary
transformation $U(t)$. The nonzero diagonal elements are
$U_{nm}(t)=\delta_{nm}\exp[i\phi_{n}(t)]$, and the phases are
$\phi_{3}=-\sin(2 \pi f t)A/hf$, $\phi_{4}=-2\pi f t$, otherwise
$\phi_{n}=0$. This transformation eliminates the time dependence
from the energy detuning and transfers it to the tunnel couplings.
It also shifts downwards by $-hf$ the bare energy level
$\Delta^{+}$ of $|T_{-}\rangle$, because near the anticrossing
point we are interested in the transitions satisfying
$\Delta^{+}-hf \approx -\Delta^{+}$, where $-\Delta^{+}$ is the
bare energy level of $|T_{+}\rangle$. The transformed Hamiltonian
is
\begin{equation}
W =\left(%
\begin{array}{ccccc}
  0 & 0 & \sqrt{2}T_{\mathrm{c}} & 0 & \Delta^{-} \\
  0 & -\Delta^{+} & a_{\mathrm{so}} & 0 & 0 \\
  \sqrt{2}T^{*}_{\mathrm{c}} & a^{*}_{\mathrm{so}} & -\varepsilon & b_{\mathrm{so}} & 0 \\
  0 & 0 &  b^{*}_{\mathrm{so}} & \Delta^{+}-hf &0\\
  \Delta^{-} & 0 & 0 & 0 & 0 \\
\end{array}%
\right),
\end{equation}
and the tunnel couplings are
\begin{equation}
\begin{split}
T_{\mathrm{c}} = -t_{\mathrm{c}}\sum^{\infty}_{m=-\infty}
(-1)^{m}J_{m} e^{i m 2\pi f t}, \\
a_{\mathrm{so}} = -t_{\mathrm{so}} \sum^{\infty}_{m=-\infty}
(-1)^{m}J_{m} e^{i m 2\pi f t}, \\
b_{\mathrm{so}} = -t_{\mathrm{so}} \sum^{\infty}_{m=-\infty} J_{m}
e^{i (m-1) 2\pi f t},
\end{split}
\end{equation}
where $J_m$ is a Bessel function of the first kind with the
argument $A/hf$ $[J_m=J_m(A/hf)]$. This Hamiltonian is exact, and
we proceed by assuming that the time independent terms of this
Hamiltonian can well describe the relevant dynamics. Thus, in the
above tunnel couplings we ignore all the time dependent terms:
\begin{equation}\label{coupls}
T_{\mathrm{c}} = -t_{\mathrm{c}} J_0, \quad a_{\mathrm{so}} =
-t_{\mathrm{so}} J_0, \quad b_{\mathrm{so}} = -t_{\mathrm{so}}
J_1,
\end{equation}
and in the transformed (moving) frame we arrive at an approximate
time independent Hamiltonian $W_0$ which has some interesting
properties.

For example, the energy spectrum of $W_0$ can be used to predict
the current resonances by examining the formation of the
anticrossing points. Especially in the regime $A\ll hf$, the
spectrum of $W_0$ approximates very well the exact Floquet
spectrum. Most importantly, the diagonalization of $W_{0}$ reveals
the existence of the eigenstate~\cite{note2}
\begin{equation}
c_+|T_{+}\rangle+c_-|T_{-}\rangle,
\end{equation}
when $\Delta^{+} = hf/2$ or, equivalently, $(g_1+g_2)
\mu_{\mathrm{B}} B = h f$, with the coefficients $c_+/c_- = -
J_1/J_0$. This eigenstate contains no $|S_{02}\rangle$ component,
which is responsible for the current; therefore, it acts as a
``dark'' eigenstate. Namely, it does not allow the ac field to
enhance the current, and consequently, the ac-induced current
($A\ne0$) is approximately equal to the background current
($A=0$). This dark eigenstate which has a Bell-like structure, is
the origin of the current antiresonance described above (e.g.
Fig.~\ref{image}). As also emphasized, the antiresonance exists
independent of the magnitude of the ac amplitude as well as for a
vanishingly small $g$-factor difference, in agreement with the
existence of the dark eigenstate predicted by $W_0$.

The predictions of the approximate Hamiltonian $W_{0}$ are
accurate enough in the regime where $\Delta^{+}$ is different from
$\varepsilon$, but when $\Delta^{+}\approx\varepsilon$ another
treatment can be sought for improved accuracy.~\cite{note3} This
observation can be understood by inspecting the exact Floquet
Hamiltonian as derived now from $W(t)$ instead of
$H_{\mathrm{DD}}(t)$. In particular, the coupling terms between
the diagonal elements $W_{ii}\pm n h f$ of the Floquet Hamiltonian
suggest that a general Floquet mode should include at least the
basis states $\exp(i n 2\pi f t)|S_{11}\rangle$, $n=0$, 1. The
coupling terms which involve Bessel functions $J_{m}$ with $|m|>1$
can typically be ignored within an approximate Floquet
Hamiltonian. Some additional properties of $W_{0}$ are examined in
Sec.~\ref{transport}.C, where the dependence of the ac-induced
current peaks on the ac amplitude is investigated.

\subsection{Current versus ac amplitude}

\begin{figure}
\includegraphics[width=5.7cm, angle=270]{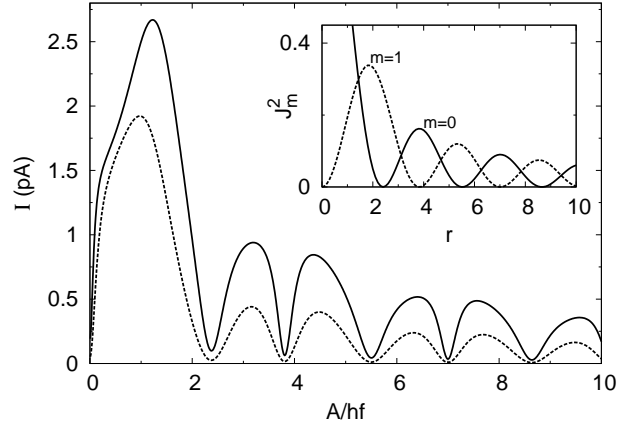}\\
\caption{Current as a function of ac amplitude $A$ at fixed
magnetic field $B=0.134$ T. For the solid line the ac frequency is
$f=26.53$ GHz and the SOI tunnel coupling is $t_{\mathrm{so}}=3$
$\mu$eV. For the dotted line $f=26.85$ GHz and
$t_{\mathrm{so}}=1.5$ $\mu$eV. In both cases the frequency
satisfies $f = (E_{5} - E_{2})/h$ as depicted in Fig.~1. The inset
shows $J^{2}_{m}(r)$, $m=0$, 1 for $0<r<10$.}\label{ivsA}
\end{figure}

\begin{figure}
\includegraphics[width=13.cm, angle=270]{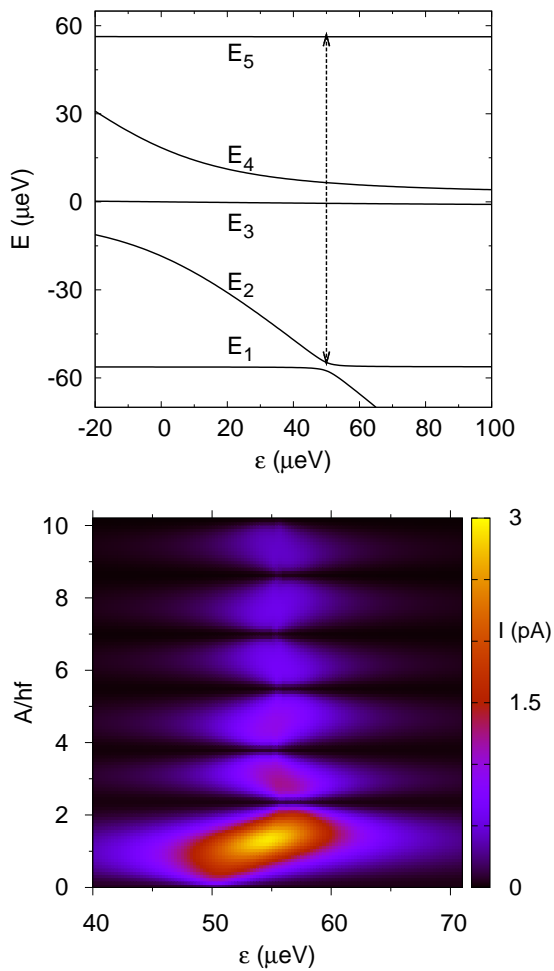}\\
\caption{The upper frame shows the two-electron eigenenergies as a
function of energy detuning for the magnetic field $B=0.134$ T.
The vertical arrow at the anticrossing point $\varepsilon \approx
50$ $\mu$eV specifies the ac frequency $f$ used in the lower
frame, specifically, $f=(E_{5}-E_{2})/h = 26.85$ GHz. The lower
frame shows the current as a function of energy detuning and ac
amplitude at fixed frequency $f=26.85$ GHz and magnetic field
$B=0.134$ T.}\label{ivsAe}
\end{figure}

According to Fig.~\ref{ivsf} the height of the current peaks
induced by the ac field depends sensitively on the ac amplitude
$A$, and exhibits a non-monotonous behaviour. Furthermore, the
approximate Hamiltonian $W_{0}$ reveals the possibility of tuning
the DD system to the so-called ``coherent destruction of
tunnelling'' regime,~\cite{flqmas2, destruct1, destruct2,
destruct3, destruct4} where the inter-dot tunnel coupling vanishes
for specific values of the ratio $A/hf$. In our ac driven DD there
are three effective tunnel coupling terms given in
Eq.~(\ref{coupls}), which are sensitive to the ratio $A/hf$. When
$J_{0}(A/hf)=0$ the spin-conserved tunnel coupling vanishes,
$T_{\mathrm{c}}=0$, and therefore $|S_{11}\rangle$ and
$|T_0\rangle$ are blocked states ($|\uparrow,\downarrow\rangle$,
$|\downarrow,\uparrow\rangle$) because they cannot coherently
tunnel to the $|S_{02}\rangle$ singlet. Simultaneously, when
$J_{0}(A/hf)=0$ the SOI spin-flipped tunnel coupling between the
$|T_{+}\rangle$ and $|S_{02}\rangle$ states vanishes because
$a_{\mathrm{so}}=0$, and thus $|T_{+}\rangle$ is also a blocked
state. In this regime, the ac-induced current should be suppressed
because only the $|T_{-}\rangle$ state is tunnel-coupled to the
$|S_{02}\rangle$ state. Similarly, the current should also be
somewhat suppressed when $J_{1}(A/hf)=0$, because the SOI tunnel
coupling between the $|T_{-}\rangle$ and $|S_{02}\rangle$ states
vanishes ($b_{\mathrm{so}}=0$) and now $|T_{-}\rangle$ acts as a
blocked state.

To study the dependence of the ac-induced current on the ac
amplitude we focus on the transition between $|\psi_{5}\rangle$
and $|\psi_{2}\rangle$, so that the ac frequency is $f =(E_{5} -
E_{2})/h$, and use the equation of motion [Eq.~(\ref{master})] to
determine the current characteristics in the steady state. In
Fig.~\ref{ivsA} we plot the current as a function of the ac
amplitude for $t_{\mathrm{so}}=1.5$ $\mu$eV and
$t_{\mathrm{so}}=3$ $\mu$eV. For these two cases the frequency $f$
is different since the SOI gap is different. For convenience, we
also plot $J^{2}_{m}$, $m=0$, 1. Some of the current
characteristics can be understood using the above arguments
regarding the formation of blocked states. For example, the
current displays a local minimum (it is suppressed) when either
$J_{0}=0$ or $J_{1}=0$. Moreover, in the asymptotic regime,
defined for $A > 2.5hf$, the current displays an oscillatory
behaviour and the overall current decreases following the overall
reduction in the interdot tunnel coupling terms
[Eq.~(\ref{coupls})] between the spin blocked states and the
$|S_{02}\rangle$ singlet state. In Fig.~6 the field corresponds to
the anticrossing $B\approx 0.134$ T, but the SOI forms another
anticrossing at $B\approx 0.014$ T (Fig.~1). At this field the
ac-induced current due to the transitions between $|\psi_2\rangle$
and $|\psi_4\rangle$ (or $|\psi_5\rangle$) has different form from
that in Fig.~6, but these transitions are not considered in the
present work.

As seen in Fig.~\ref{ivsA} for $A\lesssim 2.5hf$, two regimes can
be identified. Specifically, as the amplitude $A$ increases, the
current first increases and then it starts to decrease. The
increase of the current is expected because the ac field induces
transitions (Fig.~1) between states with different populations; a
state with mostly triplet character and high population
($|\psi_{5}\rangle$), and a state with large $|S_{02}\rangle$
component and lower population ($|\psi_{2}\rangle$). The
approximate Hamiltonian $W_{0}$ suggests that the increase of the
current is related to the increase of the tunnel coupling term
$b_{\mathrm{so}}$. This term gradually lifts the (partial) spin
blockade due to the blocked $|T_{-}\rangle$ state, by allowing
transitions from $|T_{-}\rangle$ to $|S_{02}\rangle$. However, the
tunnel couplings $a_{\mathrm{so}}$ and $T_{\mathrm{c}}$ decrease
with $A$; therefore, the current should reach a maximum value and
then should start to decrease. The crossover point is sensitive to
the exact frequency (magnetic field) and the dot-lead rate
$\Gamma$. The largest ac amplitude $A$ considered in this work is
on the order of 1.5 meV (when $A\approx 10hf$), and such
relatively large amplitude can usually be generated in quantum dot
systems by applying electrical pulses.~\cite{perge12, ono17,
bertrand, pulses1} Electrical noise in quantum dots is device
dependent and may influence the current characteristics, but
coherent effects due to the ac field have been demonstrated in
various devices when the noise level is low, enabling transport
spectroscopy of spin states and estimation of the spin-orbit
gap.~\cite{perge12, ono17}

In double quantum dots the energy detuning can usually be
controlled by gate voltages,~\cite{hanson07} and thus it is
interesting to explore the detuning dependence of the ac-induced
current near the SOI singlet-triplet anticrossing point. For
convenience in the upper frame of Fig.~\ref{ivsAe} we plot the
two-electron eigenenergies as a function of the energy detuning
($A=0$) for the magnetic field $B=0.134$ T. The anticrossing point
which is formed at $\varepsilon=0$ is due to the $|S_{11}\rangle$,
$|S_{02}\rangle$ coupling and it exists even for zero SOI. Here,
we focus on the region near the SOI singlet-triplet anticrossing
point, formed at $\varepsilon\approx 50$ $\mu$eV, and plot in the
lower frame of Fig.~\ref{ivsAe} the current as a function of the
energy detuning and ac amplitude. For all the calculations, the
Floquet-Markov equation of motion Eq.~(\ref{master}) is used
again. The field is $B=0.134$ T and the ac frequency is $f=26.85$
GHz with $f = (E_{5}-E_{2})/h$ at $\varepsilon\approx 50$ $\mu$eV.
Therefore, we refer to this particular detuning as the
``resonant'' detuning where the current is expected to be high,
whereas for this particular case under study $f\ne
(E_{5}-E_{1})/h$ at any $\varepsilon$.

In Fig.~\ref{ivsAe}, a high-current region can be identified in
the detuning range $47$ $\mu$eV $ \lesssim  \varepsilon \lesssim$
62 $\mu$eV, especially for $A/hf  \lesssim 2.5$. This range of the
detuning includes the resonant detuning value as well as the two
detuning values satisfying the resonant conditions suggested by
the approximate Hamiltonian $W_0$, i.e., $\varepsilon \approx
\Delta^{+}$ and $\varepsilon \approx hf - \Delta^{+}$ which give
$\varepsilon \approx 54.7$ $\mu$eV and $\varepsilon \approx 56.2$
$\mu$eV respectively. These two values are greater than the
resonant detuning, and as a result the extent of the high-current
region along the detuning axis is larger for $\varepsilon>50$
$\mu$eV. But, for $\varepsilon<50$ $\mu$eV the high-current region
decays faster as the system is gradually tuned off resonance. When
the ac amplitude increases for $A/hf>2.5$, the current displays
minima at the values of $A/hf$ which generate ac-induced blocked
states, $J_0=0$ or $J_1=0$, and the current pattern is similar to
that presented in Fig.~\ref{ivsA}. The minima have a
characteristic wide shape off resonance which becomes narrower
near a resonance where the current increases. The details of the
pattern of the current versus $A$ and $\varepsilon$, depend on the
choice of the exact ac frequency $f$. In Fig.~\ref{ivsAe}, the
frequency is $f = (E_{5}-E_{2})/h$ but a very similar pattern
occurs for $f = (E_{5}-E_{1})/h$, and in general for choices of
frequencies away from the anticrossing point. The high-current
regions can be easily identified by considering the corresponding
resonant conditions which involve the parameters $\varepsilon$,
$f$ and $\Delta^{+}$. In contrast, the current is in general lower
off resonance and when blocked states are formed.

\section{Conclusion}\label{conclusion}

In this work we considered a double quantum dot in the spin
blockade regime and in the presence of an ac electric field which
periodically changes the energy detuning. We focused on specific
energy configurations (Fig.~1) which involve two SOI-coupled
singlet-triplet states forming an anticrossing point, and a third
state with mostly triplet character. We studied the electronic
transport characteristics at the anticrossing point and found
strong ac-induced current peaks, in contrast to the vanishingly
small peaks observed for a pair of singlet-triplet
states.~\cite{ono17} We showed that for small ac field amplitudes
the current peaks map-out the two-electron energy levels and the
SOI-induced anticrossing point. In this case, the gap of the
anticrossing can be estimated, giving direct information about the
strength of the SOI. As the ac amplitude increases, the resonant
pattern changes drastically and a current antiresonance is formed.
Eventually, the SOI anticrossing point can no longer be probed. We
examined the ac-induced current versus the ac amplitude and showed
that current suppression can take place when the ac field gives
rise to blocked states for specific values of the ac amplitude and
ac frequency. As a result, the pattern of the current consists of
low- and high-current regions, which can be controlled by the ac
field.

The weak driving regime in which the resonant current peaks
map-out the SOI anticrossing point has been demonstrated in
different double quantum dot systems. However, the stronger
driving regime in which the resonant current peaks strongly
overlap and/or coherent interdot tunnelling is suppressed seems to
remain unexplored. In our work, we demonstrated a realistic range
of parameters for which the crossover from the weak to the strong
driving regime can be identified, and pointed out possible
experimental implications.

\setcounter{secnumdepth}{0} 

\section{Acknowledgement}

Part of this work was supported by CREST JST (JP-MJCR15N2) and by
JSPS KAKENHI (18K03479).

\setcounter{secnumdepth}{1}

\appendix

\section{Double dot in the spin blockade}

In this appendix we derive the double quantum dot Hamiltonian used
in the main article. Specifically, we employ the two-site Hubbard
Hamiltonian
\begin{equation}\label{htn}
\begin{split}
&h_{\mathrm{DD}}= \sum_{i=1}^{2}  \epsilon_{i}( n_{i\uparrow}
+ n_{i\downarrow}  ) + \sum_{i=1}^{2} U_{i} n_{i\uparrow}n_{i\downarrow}\\
&+ \frac{1}{2}\sum_{i=1}^{2} g_{i} \mu_{\mathrm{B}} B (
n_{i\downarrow} - n_{i\uparrow} )  + V n_1 n_2 + h_{\mathrm{T}},
\end{split}
\end{equation}
which allows for up to two electrons on each dot $i$ with $i=1$,
2. We define the number operator $n_{i}=n_{i\uparrow} +
n_{i\downarrow}$ with
$n_{i\sigma}=c_{i\sigma}^{\dagger}c_{i\sigma}$ for dot $i$ and
spin $\sigma=\uparrow,\downarrow$. The fermionic operator
$c_{i\sigma}^{\dagger}$ ($c_{i\sigma}$) creates (annihilates) an
electron on dot $i$ with on-site orbital energy $\epsilon_{i}$.
The Zeeman splitting on dot $i$ due to the applied magnetic field
$B$ is equal to $g_{i} \mu_{\mathrm{B}} B$, where $g_{i}$ is the
$g$-factor of dot $i$, and $\mu_{\mathrm{B}}$ is the Bohr
magneton. When two electrons occupy the same dot $i$ the intradot
Coulomb energy is $U_i$, and when two electrons occupy different
dots the interdot Coulomb energy is $V$.

The two dots are tunnel-coupled and tunnelling between the two
dots is modelled by the Hamiltonian
$h_{\mathrm{T}}=h_{\mathrm{c}}+h_{\mathrm{so}}$, with
\begin{equation}
h_{\mathrm{c}} = -
t_{\mathrm{c}}\sum_{\sigma}(c_{1\sigma}^{\dagger}c_{2\sigma}+\text{H.c.}),
\end{equation}
and
\begin{equation} h_{\mathrm{so}} = t_{\mathrm{so}}(
c_{1\uparrow}^{\dagger}c_{2\downarrow} -
c_{1\downarrow}^{\dagger}c_{2\uparrow} + \text{H.c.}).
\end{equation}
The Hamiltonian $h_{\mathrm{c}}$ describes electron tunnelling
between the two dots with coupling $t_{\mathrm{c}}$, which
measures the degree of overlap between the states localized in
different quantum dots. The Hamiltonian $h_{\mathrm{so}}$ accounts
for a Rashba-like spin-orbit interaction,~\cite{romeo, pan,
mireles} and induces interdot tunnelling via a spin-flip with
coupling $t_{\mathrm{so}}$. This coupling is inversely
proportional to the spin-orbit length which is sensitive to the
details of the double dot geometry as well as the direction of the
applied magnetic field relative to the spin-orbit
axis.~\cite{perge12, nowak, takahashi, stano} Because this
direction is device dependent, in our work we assume different
couplings $t_{\mathrm{so}}$ in the regime
$t_{\mathrm{so}}<t_{\mathrm{c}}$ which is in agreement with other
studies.~\cite{perge12, ono17, wang, chorley11} The microscopic
details of the spin-orbit interaction are not important in our
work, since the key requirement for the formation of the
ac-induced peaks studied in the main article is the nonzero
coupling $t_{\mathrm{so}}$. As shown below the Hamiltonian
$h_{\mathrm{c}}$ can only hybridize singlet states, in contrast,
the Hamiltonian $h_{\mathrm{so}}$ leads to hybridized
singlet-triplet states. Therefore, both $h_{\mathrm{c}}$ and
$h_{\mathrm{so}}$ form anticrossing points in the energy spectrum
and the degree of state hybridization is maximum in the vicinity
of these points.

So far, the double dot Hamiltonian Eq.~(\ref{htn}) is general
enough and not specific to the spin blockade regime. We now focus
on the spin blockade regime and without loss of generality, we
assume for simplicity that $U_1=U_2=U$, $V=0$, and choose for the
orbital energies
\begin{equation}
\epsilon_{1} = + \frac{U}{2} + \frac{\varepsilon}{2}, \qquad
\epsilon_{2} = - \frac{U}{2} - \frac{\varepsilon}{2},
\end{equation}
with the parameters satisfying $U \gg t_{\mathrm{c}}$,
$|\varepsilon|$. Specifically, the charging energy $U$ can be as
large as 10-20 meV, whereas $t_{\mathrm{c}}$ is typically less
than 1 meV. The energy detuning $\varepsilon$ is usually tunable
with electrostatic gates, and quantifies the energy difference,
\begin{equation}
\varepsilon = E(1,1) - E(0,2).
\end{equation}
The notation $E(n,m)$ denotes the energy of the bare charge state
with $n$ ($m$) electrons on dot 1 (dot 2). The single electron
states of the Hilbert space are $c^{\dagger}_{i\sigma}|0\rangle$
with $i=1$, 2, spin $\sigma=\uparrow, \downarrow $ and $|0\rangle$
is the vacuum state. Because of the small ratio
$t_{\mathrm{c}}/(\epsilon_{1}-\epsilon_{2})$, the hybridization
between dot 1 states and dot 2 states is typically very small.

The two-electron states of the Hilbert space are:
\begin{equation}
\begin{split}
&|S_{20}\rangle =
c^{\dagger}_{1\uparrow}c^{\dagger}_{1\downarrow}|0\rangle, \quad
|T_+\rangle =
c^{\dagger}_{1\uparrow}c^{\dagger}_{2\uparrow}|0\rangle,\\
&|\uparrow,\downarrow\rangle =
c^{\dagger}_{1\uparrow}c^{\dagger}_{2\downarrow}|0\rangle, \quad
|\downarrow,\uparrow\rangle = c^{\dagger}_{1\downarrow}
c^{\dagger}_{2\uparrow}|0\rangle,\\
&|T_-\rangle = c^{\dagger}_{1\downarrow}
c^{\dagger}_{2\downarrow}|0\rangle, \quad |S_{02}\rangle =
c^{\dagger}_{2\uparrow} c^{\dagger}_{2\downarrow}|0\rangle.
\end{split}
\end{equation}
Alternatively, we can define states with definite spin number,
i.e., singlet states: $|S_{20}\rangle$,
$|S_{11}\rangle=(|\uparrow,\downarrow\rangle-|\downarrow,\uparrow\rangle)
/\sqrt{2}$, $|S_{02}\rangle$ and triplet states: $|T_-\rangle$,
$|T_{0}\rangle=(|\uparrow,\downarrow\rangle+|\downarrow,\uparrow\rangle)
/\sqrt{2}$, $|T_+\rangle$. The energy of the state
$|S_{20}\rangle$, e.g., when two electrons occupy dot 1, is
$E(2,0) = 2\epsilon_{1}+U = 2 U + \varepsilon$, whereas the energy
of all states with one electron on each dot is
$E(1,1)=\epsilon_{1}+\epsilon_{2}=0$, and the energy of the
$|S_{02}\rangle$ state is $E(0,2)=2\epsilon_{2}+U=-\varepsilon$.
Because of the large energy scale difference (on the order of
$2U$) the state $|S_{20}\rangle$ has a minor effect on the spin
blockade physics and can be ignored. As a result, there are five
relevant two-electron states in the spin blockade regime:
\begin{equation}
|S_{11}\rangle, \quad |T_{+}\rangle, \quad |S_{02}\rangle, \quad
|T_-\rangle, \quad |T_{0}\rangle.
\end{equation}
At low temperatures (0.1 K) three- and four-electron states are
not involved in the transport cycle and can be ignored.
Furthermore, when the Fermi energy of the right lead
$E_{\mathrm{F}}$ satisfies
$\epsilon_2<E_{\mathrm{F}}<\epsilon_2+U$, a single electron
occupies dot 2 during the transport cycle while a second electron
is allowed to tunnel from dot 2 to the right lead.

To account for the effect of the ac electric field we assume that
the orbital energies of the two dots are modulated in a
``symmetric way", thus,
\begin{equation}\label{acfield}
\begin{split}
&\epsilon_{1} = + \frac{U}{2} + \frac{\varepsilon}{2} -
\frac{A}{2}\cos(2\pi f t),\\
&\epsilon_{2} = - \frac{U}{2} - \frac{\varepsilon}{2} +
\frac{A}{2}\cos( 2 \pi f t).
\end{split}
\end{equation}
The amplitude of the ac field is $A$ and the frequency is $f$. The
assumption of symmetric modulation is not unique; we can
equivalently assume, for instance, that $\epsilon_{1}$ is
unaffected by the ac field and $\epsilon_{2} = - \frac{U}{2} -
\frac{\varepsilon}{2} + A \cos( 2 \pi f t)$. In this case the
conclusions in the main article remain unchanged. The particular
choice in Eq.~(\ref{acfield}) allows us to define the time
dependent energy detuning
\begin{equation}
\delta = - \varepsilon + A\cos( 2 \pi f t),
\end{equation}
and write the orbital energies as
\begin{equation}
\epsilon_{1} = + \frac{U}{2} - \frac{\delta}{2}, \quad
\epsilon_{2} = - \frac{U}{2} + \frac{\delta}{2}.
\end{equation}
Then, in the two-electron basis $|S_{11}\rangle$, $|T_+\rangle$,
$|S_{02}\rangle$, $|T_-\rangle$, $|T_0\rangle$, the Hamiltonian
$h_{\mathrm{DD}}$ has the form given in Eq.~(1) in the main
article. This approximate Hamiltonian is valid in the spin
blockade regime and has also been employed in other
works.~\cite{perge12, giavaras19, pei, giavarasE} From this
Hamiltonian we see that the only time dependent term corresponds
to the energy of the $|S_{02}\rangle$ singlet state (diagonal
term) and is equal to the detuning $\delta$. Even though, the
Coulomb energy $U$ is the largest energy scale, it does not
explicitly appear in the Hamiltonian because the $|S_{20}\rangle$
state is ignored. The Hamiltonian Eq.~(1) in the main article also
shows that when $A=0$ the $|S_{02}\rangle$, $|S_{11}\rangle$
states are hybridized due to the $t_{\mathrm{c}}$ term, and
similarly the $|S_{02}\rangle$, $|T_{\pm}\rangle$ states are
hybridized due to the $t_{\mathrm{so}}$ term. In particular, the
ac-induced current peaks studied in the main article are formed
only when $t_{\mathrm{so}}\ne 0$.



\begin{thebibliography}{10}
\bibitem{note2}
For simplicity, we use the same notation for the basis states in
the moving and laboratory frame.
\bibitem{note3}
Another transformation $U^{'}(t)$ can be performed with the
non-zero phases: $\phi_{3}=-\sin(2 \pi f t)A/hf$,
$\phi_{1}=\phi_{4}=\phi_{5}=-2\pi f t $; leading to
$T_{\mathrm{c}} = -t_{\mathrm{c}} J_1$ in Eq.~(12) whereas
$a_{\mathrm{so}}$, $b_{\mathrm{so}}$ are the same as in Eq.~(12).
An approximate calculation of the current using both $U(t)$ and
$U^{'}(t)$ (when $\Delta^{+}\approx\varepsilon$ and near the
anticrossing) can be in better agreement with the exact current
than using only $U(t)$. This approximate calculation is not
pursued in this work.
\end{thebibliography}
\end{document}